# Near-Term Self-replicating Probes - A Concept Design

Olivia Borgue[1], Andreas M. Hein[1],

[1]Initiative for Interstellar Studies (i4is), 27-29 South Lambeth Road, London SW8 1SZ, United Kingdom,
borgue@chalmers.se, andreas.hein@i4is.org

**Abstract**

Self-replicating probes are spacecraft with the capacity to create copies of themselves. Self-replication would potentially allow for an exponential increase in the number of probes and thereby drastically improve the efficiency of space exploration. Despite this potential, an integrated assessment of self-replicating space probes has not been presented since the 1980s, and it is still unclear how far they are feasible. In this paper, we propose a concept for a partially self-replicating probe for space exploration based on current and near-term technologies, with a focus on small spacecraft. The purpose is to chart a path towards self-replication with near-term benefits, rather than attempting full self-replication. For this reason, components such as microchips and other microelectronic components are brought with the initial probe and are not replicated. We estimate that such a probe would be capable of replicating 70% of its mass. To further increase this percentage, we identify technology gaps that are promising to address. We conclude that small-scale, partially self-replicating probes are feasible near-term. Their benefits would play out in exploration missions requiring roughly more than a dozen of probes.

## 1. Introduction

A self-replicating probe is a spacecraft with the capacity to create copies of itself for exploring space [1]. It creates copies by using space resources that are available and accessible, such as on planets, moons, or asteroids [1,2]. While self-replicating probes have been primarily proposed for space exploration [3], the underlying technology might also be used for constructing space infrastructure [4] and settling the universe [5]. The main advantage of self-replication is exponential growth in the number of spacecraft, which would enable rapid exploration of space or rapid bootstrapping of space infrastructure. Concepts for self-replicating probes have been proposed in the literature for decades, but they remain hypothetical to date [3,4,6,7].

A proof of concept of a self-replicating probe might have significant consequences. Such probes might allow for the exploration of every corner of the Milky Way and even beyond in short astronomically and evolutionary time scales [5,8,9]. According to some estimates, it would take a self-replicating probe approximately half a million years to manufacture millions of probes across the Milky Way, assuming each one travels at approximately 1/10th the speed of light [8].

The theory of self-replicating probes is based on von Neumann's theory of self-replicating machines [10], which is the reason why they are often called von Neumann probes. Although the existence of a self-replicating machine has been formally proven [11], an actual construction, even in the form of a computer program, is difficult and computationally expensive [12,13]. Hence, challenges of developing self-replicating probes are not only encountered at the level of the hardware but also the software.

Currently, the only self-replicating hardware is self-assembling systems [13]. These systems are collections of passive elements that self-assemble under external agitation or naturally occurring physical forces. Such self-reconfigurable robots have been investigated by a number of authors, which were previously reviewed by Chirikjian [13] and Toth-Fejel et al. [14]. In most cases, modular components are all connected (either physically or by a communications link), and the topology of that connection changes as a function of time or the task requirements. Such is the case of the self-replicating systems proposed by Penrose [15] and machines comprising four different components that are assembled by following tracks [11]. Further robotic self-replicating machines have been proposed but similarly use prefabricated parts that are assembled to form copies of themselves [16–19]. Several NASA NIAC studies [14,20–22] have concluded that at least "cranking" self-replicating machines are feasible.

Nevertheless, for any practically useful application, it has been argued that physical self-replicating machines would need to possess considerable computing power and significant manufacturing capabilities, involving a whole self-replication infrastructure [1,3,14,23]. For example, to develop a fully self-replicating probe, the development of generic mining and manufacturing processes, applicable to replicating a wide range of components, and automation of individual steps in the replication process as well as supply chain coordination, are necessary. A possible solution to this challenge is partial self-replication, where complete self-replication is achieved gradually when the infrastructure is built up [6]. Thus, the remaining engineering challenges are still considerable.

Concepts for self-replicating probes were proposed by authors such as Bond and Martin [24], for the Daedalus interstellar probe or Freitas [3], for the REPRO concept. These authors proposed large (hundreds to thousands of tons) self-replicating probes for interstellar exploration, which would use gas giants and moons for the establishment of resource harvesting and self-replication facilities. However, they do not provide a concrete system architecture to demonstrate the feasibility of their concepts. Chirikjian [13] and Boston et al. [20] proposed concepts and architectures for self-replicating systems; however, those are not envisioned to be developed with current technologies. Langford et al. [26] propose a hierarchical self-replicating spacecraft concept, focusing on defining standardized, modular spacecraft parts, fitting into a 3U-CubeSat. The use of in-situ resources and the concept of operations are not considered in this bottom-up approach. Jones [27], as well as Hein and Baxter [1], reviewed the feasibility of self-replicating probes concluding that the main challenges are related to low computing power and the maturity of artificial intelligence applications. No concept of a self-replicating system able to be manufactured with the current state of the art technologies is proposed.

It seems that to date, no feasibility assessment of self-replicating spacecraft has been conducted, based on current and near-future technologies. The advent of novel manufacturing technologies such as 3D-printing / additive manufacturing increased the possibility to manufacture a wide range of components, with the potential to drastically reduce the necessary infrastructure for replicating the probe. Furthermore, the miniaturization of spacecraft opens the possibility to significantly reduce the size and mass of spacecraft, reducing the required manufacturing capacities for replication.

Therefore, the objective of this paper is to propose a concept for a self-replicating probe for space exploration based on current and near-term technologies, with a focus on small spacecraft (here: < 100 kg). Instead of a bottom-up approach, such as in Langford et al. [26], we use a top-down approach, starting from mission objectives, defining system modules and technology alternatives, and proposing a concept design for a self-replicating probe. Instead of aiming at complete replication, we estimate the degree of self-replication, which is achievable with current and near-term technologies. We will also provide estimates for the economic breakeven point for such probes, compared to non-replicating probes.

In Section 2, we present the mission and system drivers, in Section 3, technology alternatives and their assessment, and in Section 4, we present our concept for a near-term self-replicating probe. This is followed by a cost comparison between using such probes and conventional probes for an exploration mission. Section 5 presents a technology roadmap towards full self-replication, miniaturization, and the exploration of other star systems.

## 2. Mission and system drivers

The focus of this article is to propose a concept design of a von Neumann probe, manufactured with current or near-term state of the art technologies, able to (partially) self-replicate with available materials from celestial bodies. In the following, we first establish generic objectives and define high-level requirements for such a probe, without constraining the operational domain, e.g. cis-lunar space, inner / outer solar system, interstellar. The technology analysis in Section 4 will then impose constraints on feasible operational domains, based on the technological baseline.

In the following, it is assumed that the main purpose of the probe is space exploration. Hence, the probe must fulfill its self-replicating functions and collect and transmit data back to Earth. To perform these specified functions, the system architecture has the hereby introduced functional requirements:

**Self-replication**

- Harvest material from celestial bodies.
- Manufacture its systems with the harvested resources
- Generate power to feed the self-replication associated systems
- Land on celestial bodies for resource harvesting
- Control the self-replication associated systems
- Navigate, take-off and land

**Data collection and transmission**

- Generate power to feed the data collection and transmission systems associated systems
- Control the systems associated with self-replication
- Gather data from surroundings
- Transmit data back to Earth

Considering the stated system requirements, such a probe would be composed by a system architecture as the one presented in Figure 1.

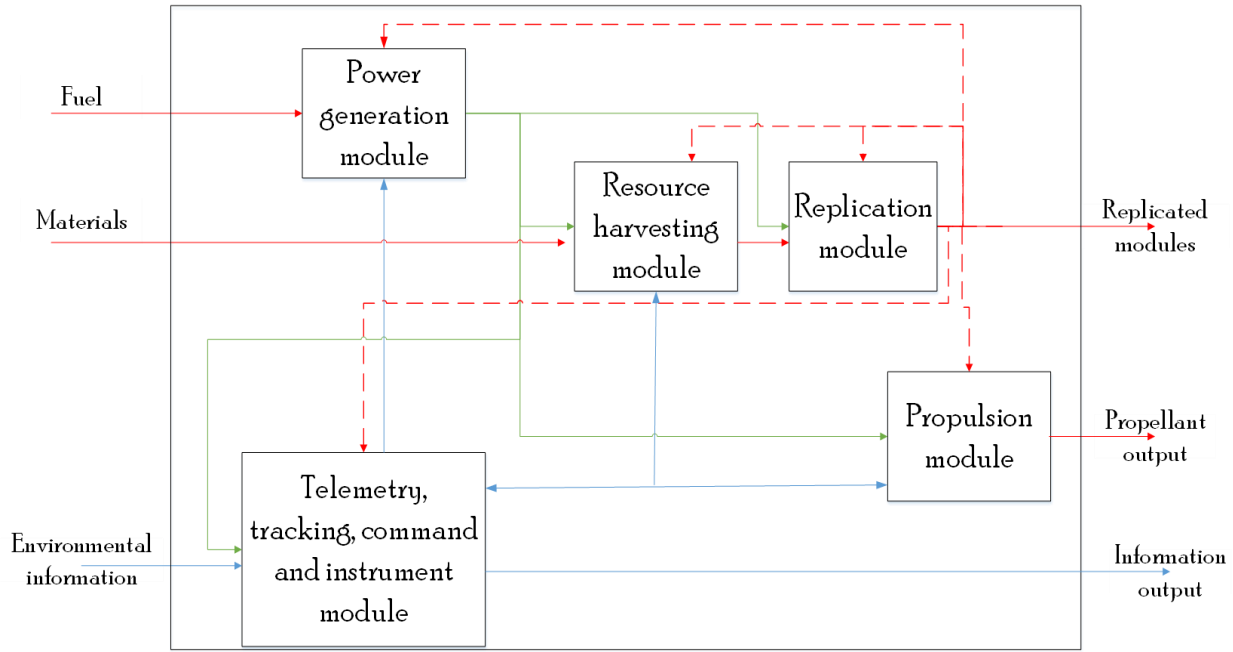

**Figure 1. System architecture for a self-replicating probe.**

The probe would encompass six different modules: Power generation, resource harvesting, replication, propulsion, control, and telemetry, tracking, command and instrument module. The system inputs include fuel, materials for self-replication and environmental information. The outputs are information transmitted to Earth, propellant waste, and replicated modules. The dashed red lines connecting the replication module with the other modules contemplate the possibility of implementing the replication module for repairing possible malfunctions on the other modules, a desirable function for interstellar exploration [2].

The system requirements and architecture previously presented describe the ideal capabilities of a self-replicating probe for interstellar exploration. In this article, however, the technological capabilities of current and near-term technologies required to manufacture each module are analyzed. A discussion about how the current state-of the art technologies restrict the operational capabilities of a self-replicating probe is provided in the final sections of this article.

# 3. Technology analysis for a self-replicating probe

In this section, various technologies for the different modules of a self-replicating probe are assesses according to the main system functions stated in Section 3.

## 3.1. Telemetry, tracking, command and instrument module

Developing a self-replicating probe with the capabilities described by Bond and Martin [24] or Freitas [3], requires the development and control of massive and sophisticated resource harvesting and replication modules. For this purpose, the probe should possess considerable computing power. However, as stated by Jones [27] the main limitation for a self-replicating probe is the TRL of current AI systems. Current AI systems would not be able to control and manage large scale self-replication endeavors including, multi-technological, independent manufacturing and harvesting facilities such as those proposed by Freitas [3]. Moreover, AI systems are also in their infancy regarding autonomous navigation, attitude control and landing systems. To perform these tasks, a concept like an artificial general intelligent (AGI) system must be developed [1].

For these reasons, regarding telemetry and telecommand, a feasible self-replication probe would have a certain degree of autonomy but some of its functions (such as some navigation features) would still be managed from Earth.

As established in section 3, a self-replicating probe must be able to control the self-replication associated systems, gather data from surroundings and transmit data back to Earth. In that context, a guidance, navigation and control (GNC) system is necessary, which can be considered as a combination of an orbit determination and control subsystem (ODCS) and an attitude determination and control subsystem (ADCS). For an ADCS, sensors such as star trackers, sun sensors, or magnetometers are necessary to determine spacecraft attitude and actuators such as magnetorquers, reaction wheels, or thrusters to orient the probe [28,29].

A command and data handling system must be connected to all on-board systems to process and store and transmit data in the probe, as well as real-time and historical telemetry, and full probe monitoring and control.

Hodges et al. [30] provide a comprehensive assessment of antenna options for the first CubeSat for deep space, the MarCO mission. For the MarCO mission, a novel deployable reflectarray was preferred due to the reduced available area in a CubeSat. This alternative, however, is also convenient for a self-replicating probe where the different antenna modules can be manufactured separately and then assembled, reducing the size of the replication module. Deployable mesh reflectors could also be manufactured in a reduced volume and then assembled; however, they present a higher deployment complexity and, therefore, a lower reliability. Deployment complexity problems are eliminated with the implementation of no-deployable antennas, such as patch array antennas [30] or the parabolic antennas implemented for Cassini, Voyager or Osiris-Rex [31–33]. However, the large size (between 2m and 4 2m) of those parabolic antennas can render their manufacturing process in the probe difficult. In Table 1, a comparison of different antenna alternatives for a self-replication probe is presented.

**Table 1. Antenna comparison [29, 31].**

|  | **Reflectarray** | **Mesh reflector** | **Non-deployable, parabolic** | **Non-deployable, patch array** |
|---|---|---|---|---|
| Gain | High gain no restricted to manufacturing volume | | Restricted to manufacturing volume | |
| Deployment complexity | Low | High | None | |
| Reliability | High | Medium | High | |
| Mass | Low | Low | High | Low |

In an arrangement similar to the one implemented for the MarCO mission, a space of a 6U CubeSat can house batteries, flight computer and attitude control system, radio and navigation equipment and thrusters.

For convenience, we consider the instruments, which are used for gathering data of celestial bodies, as part of this module. The reason is that the components of the telemetry, tracking, and command module and instruments consist of electronic and high-precision optomechanical components. Analysis of celestial body surfaces can be performed by techniques such as photoelectric photometry [34], radiometry [35], polarimetry and spectropolarimetry [36], hyperspectral imaging [37], thermal modelling [38], etc. Miniaturized equipment for these purposes was implemented in the design of the NEA Scout, Lunar Flashlight, Lunar IceCube or LunaH-Ma CubeSats [28].

## 3.2.  Replication module

According to literature [1,39], a convenient manufacturing technology for a self-replication probe is additive manufacturing. In additive manufacturing technologies, material is added layer by layer to form a 3D geometry, based on a computer file. As resources in space might be scarce or difficult to find, additive manufacturing technologies can be advantageous for reducing material waste and the possibility to manufacture intricate geometries [40,41]. In Table 2, a comparison of different additive manufacturing technologies and their evaluation in the context of a self-replicating probe is presented.

As most of the material obtained from harvesting procedures would most likely be powder (refer to Section 4.4), laser powder bed (LPB) additive manufacturing technologies could be an attractive alternative, as they would require reduced material processing. Moreover, laser-based additive manufacturing can enable the manufacturing of both metals and thermoplastic polymers components in space, with little post-processing needed [42].

Additive manufacturing has previously attracted the attention of the Asteroid Mining Corporation [43] that proposed the implementation of additive manufacturing technologies for their Asteroid Mining Probe One (AMP-1) 2028. The company Planetary Resources has also developed a proof of concept of an additive manufactured object made of meteorite powder based on Iron, Nickel and Cobalt [44]. In this context, a couple of mixed metals additive manufacturing machines have also been developed by other parties such as A222 from Formalloy [45] and NVLABS [46]. Both implementing metal deposition.

Moreover, there is an additive manufacturing facility aboard the ISS since 2016, were plastic components are manufactured with a Fused Deposition Modeling (FDM) machine developed by Made in Space Inc [42]. On the same line, the Chinese Academy of Sciences developed an FDM 3D machine for µ-gravity and a digital light processing additive manufacturing technology for producing ceramic green bodies [46]. A laser powder bed additive manufacturing machine for µ-gravity has also been developed by Zocca et al. [42], proposing a gas flow-assisted powder deposition additive manufacturing system that keeps the powder against the building platform, reducing the risk of having metal powder unrestrictedly suspended in the additive manufacturing machine.

**Table 2. Comparison of additive manufacturing technologies [41,47]**

| Additive manufacturing technology | Materials | Benefits | Limitations | Applications |
|---|---|---|---|---|
| Material extrusion | Polymer, metal, composites, biomaterials, | Small machines. Multimaterial. Easy to replicate. Large building volume. Low energy consumption. | Highly anisotropic, shrinkage, roughness. Needs support | Structural components, propulsion, electronics |
| Material Jetting | Polymer, wax, metals | High resolution. Multimaterial. Good surface quality | Difficult to replicate. Large | Solar cells, electronics |
| Binder jetting | Metal, organic materials, ceramics, glass, sand | Multimaterial. High TRL | Large, Shrinkage, anisotropic. Powder, reduced building volume | Structural components, propulsion |
| Vat photopolymerization | Ceramics, photopolymer resin | Requires curing. Good surface quality. | Liquid resin. Difficult to replicate | Electronics |
| Powder bed fusion | Polymers, metals | Strong and durable parts. High TRL. Multimaterial. | Large, Powder, reduced building volume. High energy consumption. | Structural components, propulsion, tools, electronics |
| Direct energy deposition | Metals | Strong and durable parts. High TRL. Multimaterial. Large building volume. | Large, Powder, difficult to replicate. High energy consumption. | Repairment, Structural components, propulsion, tools |
| Laser chemical vapor deposition | Metalorganics, metals, ceramics, composites | High precision, good surface quality. High energy consumption. | Large, difficult to replicate | Solar cells, electronics |

Some problems that still arise with LBP additive manufacturing systems are related with a high energy consumption, chemical compatibility differences in melting points of the material collected and non-uniform powder size. Moreover, LPB additive manufacturing machines have a building space confined to the building chamber and their replication characteristics are poor when replicating a machine for µ-gravity.

Metal extrusion systems, however, are another viable option in which a filament, consisting of a polymer or organic compound (binder) filled with metal powder is extruded to form composite metal/binder parts that can be sintered to burn off the binder and fuse the metallic powder. The smallest variants of these systems are lightweight (<20kg) and have a low energy consumption [47].

Regarding additive manufacturing of solar cells and electronic components in general, currently available additive manufacturing systems are complex and heavy (>100kg), requiring therefore, large amounts of harvested materials and highly advanced and precise additive manufacturing technologies for their replication [48].

A more detailed assessment of the manufacturability of the different modules in the probe are presented in the following sections.

### 3.3. Power generation module

Conventional satellite systems have used solar panels for power generation, up to a distance of Jupiter. Solar panels for satellite systems can be manufactured as silicon cells covered in glass, multi-junction cells made from gallium arsenide (GaAs) and other similar materials, or perovskite-based cells [48]. The perovskite-based solar cells have been proven more efficient in laboratory environments and could be a better option for reduced physical spaces or constrained material availability [49]. As Table 4 in Section 4.4 indicates, every type of solar arrays could be theoretically considered for the probe as their based materials are present in asteroids. Moreover, currently available additive manufacturing technologies can manufacture such components [48,50,51].

Heading away from the sun, however, the solar radiation decreases from 1,374 Watts/m² around Earth, to 50 Watts/m² near Jupiter. Therefore, power generation systems based on solar energy are nearly useless outside the solar system. In this case, radioisotope thermoelectric generators (RTG) are preferred, as earlier implemented by probes such as Voyager, Cassini or Ulysses.

The Voyager probes, for instance, have three RTGs with plutonium 238 as a fuel source. As the isotope decays, it produces heat which is converted to electrical energy. Currently, advanced Stirling radioisotope generators (ASRG) are under development in NASA. ASRG are radioisotope power systems that implement a Stirling power conversion system to convert radioactive-decay heat into electricity. It is estimated that ASRGs can be as about four times more efficient previous RTGs [52].

Hydrogen-oxygen fuel cells are another option which have been implemented on missions such as Apollo or Gemini, however, under normal operation fuel cells may rapidly deplete their fuel supply [53,54].

The last power generation alternative are nuclear fission reactors [55], which make use of materials such as enriched Uranium and Thorium [56,57]. These materials, and those needed for RTG systems, were found in the surface of the moon [58,59] and Mars [60] and in meteorite samples [61]. However, the amount of those materials in asteroids between 0.1 and 0.35 p.p.m [62]. Moreover, additional equipment would be needed for performing uranium enrichment for manufacturing high-quality solid fuel.

The scarce availability of fuel for RTG/SRG or nuclear fission alternatives pose a problem to maintain the energy requirements of additive manufacturing technologies (at least >50W [63,64]). Larger proportions of U and Th seem to be present in the moons and planets such as in the Earth´s Moon (average of 2 ppm for U and 1.2 for Th [59]) and Mars (average of 5 ppm for U and 1.1 for Th [60]. However, large amounts of energy, propellant and coordination efforts are required for landing and taking off from those celestial bodies.

In Table 3, a condensed comparison of the described PGM alternatives is presented.

**Table 3. Comparison of different alternatives for the power generation module [65]**

| PGM | System alternatives | Deep space availability | Self - replicability | Other limitations |
|---|---|---|---|---|
| Solar | Solar panels | available | replicable | Sunlight |
| RTG.ASRG | Po, Sr, Pu, Am | low | replicable | May damage equipment |
| Hydrogen-oxygen fuel cells | Hydrogen-Oxygen | available | replicable | Low autonomy |
| Nuclear fission | U,Th | low | replicable* | May damage equipment |

*requires higher processing

### 3.4. Resource harvesting module

Projects such as REPRO [3] and Daedalus [24] were based on resource harvesting from Jovian planets with the creation of large replication factories. This argument is based on the large size and materials requirements of such a spacecraft which render asteroid mining insufficient. As discussed in the previous section, however, resource harvesting from large moons and planets would require large amounts of energy and propellant. Moreover, landing on large celestial bodies would require larger efforts in terms of structural and material requirements, precision landing and atmosphere/surface assessment [66,67]. For these reasons, the resource harvesting module for a state-of-the-art self-replicating probe will be based on asteroid mining strategies.

**Asteroids and resource availability**

A self-replicating probe must be able to manufacture all its components from materials available in different asteroids types. Hereby, a summary of asteroids types and their content is presented [34,68–70]:

- Dark C, carbonaceous asteroids. Most common type of asteroid, believed to be close to the Sun's composition, with little hydrogen or helium and water and other "volatile" and carbonaceous compounds.
- Bright S, siliceous asteroids. Mostly located in the inner part of the Main Asteroid Belt, closer to Mars. Made mostly out of stony materials such as metallic iron with some silicates, they are believed to be the source of most of chondrite meteorites. Their most common minerals include anorthite, melilite, perovskite, aluminous spinel, hibonite, calcic pyroxene, and forsterite-rich olivine.
- Bright M, metallic asteroids. Found in the middle region of the asteroid belt, mostly made up of metallic iron and traces of silicates.
- Other types: D type (Trojan asteroids of Jupiter), dark and carbonaceous. V type, distant asteroids between the orbits of Jupiter and Uranus, which may have originated in the Kuiper Belt. Probably made out of organic-rich silicates, carbon and anhydrous silicates, possibly with water ice in their interiors.

Authors such as Ross [71], Erickson [72] or Hellgren [73] outlined the use of different asteroid materials and their possible uses in spacecraft, as presented in Table 4:

**Table 4. Asteroids materials and their possible implementation.**

| Component | Primary use |
|---|---|
| $H_2O$, $N_2$, $O_2$ | Life Support |
| $H_2$, $O_2$, $CH_4$, $CH_3OH$ | Propellant |
| $H_2O_2$ | Oxidizer |
| $SO_2$ | Refrigeration |
| $CO$, $H_2S$, $Ni(CO)_4$, $Fe(CO)_5$, $H_2SO_4$, $SO_3$ | Metallurgy |
| Fe, Ni | Construction and manufacturing |
| Si, Al, P, Ga, Ge, Cd, Cu, As, Se, In, Sb, Te | Semiconductors |
| Au, Pt, Pd, Os, Ir, Rh, Ru, Re, Ge | Elctronics |

The technology needed to carry out asteroid mining is still underdeveloped. In the literature, however, several strategies for material harvesting in zero-gravity environments, which are similar to those implemented on Earth, are mentioned: drilling, blasting, cutting and crushing [73]. Hellgren [73], proposed the theoretical utilization of different resource harvesting methods depending on the asteroid type, presented in Table 5:

Table 5. Asteroid mining techniques depending on asteroids content [73]

| Asteroid type | Mining | Processing |
|---|---|---|
| Ice mixtures | Blast, heat, distill | Phase separation |
| Friable rock | Blast, rip | Phase separation, mech, chem, mag |
| Hard rock | Blast, disc cutters | mech, chem, mag |
| Metallic Ni-Fe | Concurrent with processing | Smelting, carbonyl methods |
| Hard rock-metallic Ni-Fe | Blast, heat, rip | mech, chem, mag; smelting |

There are currently several asteroid mining companies such as Asteroid Mining Corporation, TransAstra, Deep Space Resources or Planetary Industries [74]. These companies are still in their infancy and have yet not fully developed and tested resource harvesting technologies able to operate in space. The Optical Mining concept, from TransAstra, for instance, implements a highly concentrated sunbeam for drilling holes or excavating asteroids while they are enclosed in containment bags [75]. As the concept is based on sunlight, its implementation is only feasible within the inner solar system.

Planetary resources proposed a method for water extraction [76] where an asteroid is enclosed in a resistant bag, then heated to extract water vapor and finally released. This water extraction technique is, however, very energy-consuming.

In parallel to the theoretical concept development and fundamental science, the TRL of the critical technologies for In-Situ Resource Utilization (ISRU) is increasing. Many ISRU concepts have been demonstrated as hardware in laboratories or with terrestrial demonstrators [55].

The authors Zacny et al. [77], for instance, proposed and demonstrated an ISRU concept based on soil mining with a deep fluted auger for water extraction. Water is extracted within the flutes and the soil is discarded. Drilling in icy soil and ice was demonstrated in vacuum chambers by the authors. On a later work, the authors developed the "Sniffer" concept that is being developed to reach TRL 5 via NASA funding [77]. The "Sniffer" implements a heater deep flute with perforated walls for melting and/or sublimating volatiles.

Regarding space probes, two missions have successfully collected significant amounts of materials from asteroids, Japan's Hayabusa 2 [78] sent to asteroid Ryugu, and NASA´s Osiris-Rex [33]. Hayabusa 2 is the first spacecraft to fire a projectile and collect underground asteroids samples from the blast. Orisis-Rex implemented a high-pressure nitrogen based Touch-and-Go Sample Acquisition Mechanism (TAGSAM) that consists of a sampler head and an articulated positioning arm.

## 3.5. Propulsion module

Extensive literature reviews about propulsion methods for space applications were performed by authors such as Forward [79], Mueller [80], Larangot et al. [81] and Rossi [82]. These propulsion systems provide ΔV for orbital changes, trajectory injections, performing orientation, takeoff and landing maneuvers. A summary of their findings and a reflection about their implementation in self-replicating probes is presented in Table 6.

As previously implemented in missions such as MESSENGER, Juno or Ulysses, propulsion systems can be utilized for orientation purposes, while gravity assist maneuvers provide significant additional $\Delta V$ [83]. The main constraint for gravity assist maneuvers, however, is that planets and other celestial bodies need to be aligned to enable a maneuver to a particular destination.

An alternative propulsion method is the use of solar sails, large reflective sails that propel a spacecraft when photons collide with the sail, capturing the momentum of light from the Sun [84]. Solar sails have already been launched by the Japanese Aerospace Exploration Agency (JAXA) in the IKAROS spacecraft, by NASA in the NanoSail-D spacecraft, and by The Planetary Society with the LightSail 1 spacecraft [85]. Solar sails could be theoretically replicated by the probe; however, their performance depends on the availability of sunlight / starlight.

**Table 6. Comparison of different micro-propulsion alternatives for the propulsion module [80–82]**

| Propulsion system | Propellant alternatives | Deep space availability | Self-replicability |
|---|---|---|---|
| Cold gas | Nitrogen | Available | Difficulty to pressurize propellant |
| Bi propellant | Liquid oxygen +ethanol | Not available | - |
| Monopropellant | Hydrazine | Available | Difficulty to pressurize propellant |
| Monopropellant | Hydrogen peroxide | Available | Difficulty to pressurize propellant |
| Solid (non restartable) | Fuel+oxidizer+binder | Available | Difficulty to pressurize propellant |
| Field emission | Cesium, indium | Not available | - |
| Plasma pulsed | Teflon | Not available | - |
| Ion thruster (PPT, HT, GIT) | Xe, Teflon (PPT), bismuth, iodine, argon, krypton. Nitrogen, oxigen | Most materials are not available. | Difficulty to pressurize propellant |
| Laser ablation plasma [86] | PVC, Kapton. Any solid [87] | Available | Replicable |
| Vaporizing liquid thruster | Water [88] | Available | Replicable |
| Electrothermal (resistojets) | Hydrazine, water | Available | Replicable (water) |

In Figure 2, a comparison of the $\Delta V$ provided by the different propulsion systems performed by Rossi [82], is presented.

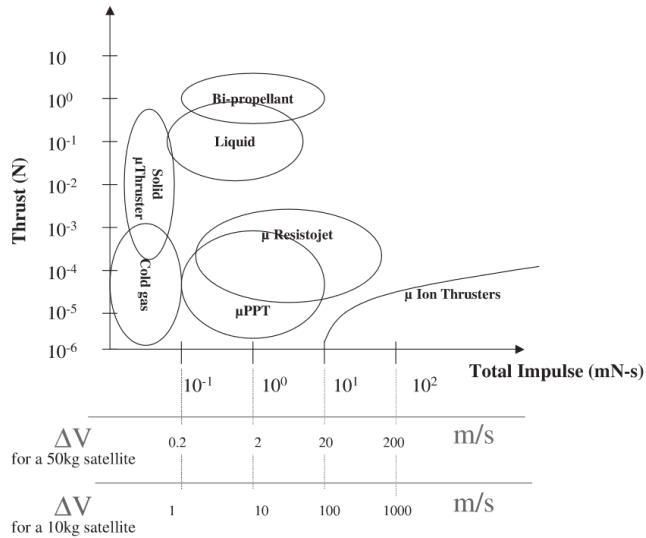

**Figure 2. ΔV comparisons of various micro-propulsion systems. Extracted from Rossi [82].**

## 4. Concept design of a self-replicating probe

Based on the technology analysis presented in Section 4, a simplified function model built from the enhanced function-means technique, EF-M [89] of a self-replicating probe is presented in Figure 3. The model represents a hierarchical array of functions, in white (such as "Store energy") and design solutions, in grey (ways to fulfill the functions, such as "Batteries"). The design space of the design solutions is limited by constraints, in black (such as "Temperature and radiation"). The different design solutions interact through "interact with" connections, represented with dotted lines. These interactions can be spatial (black), through material (red), energy (green) or information transfer (blue).

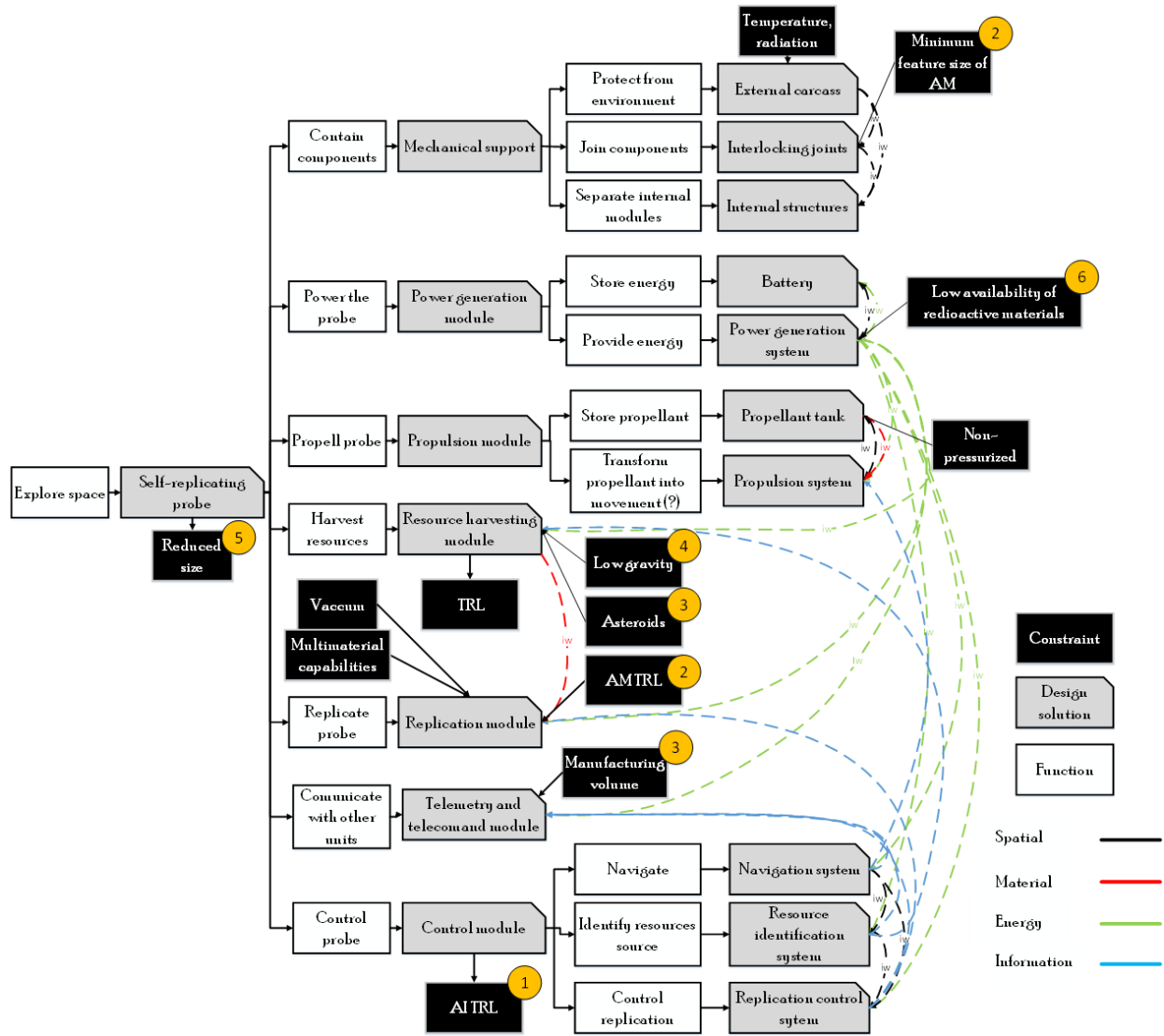

**Figure 3. EF-M model of a self-replicating probe.**

The most prominent constraints affecting the design of a self-replicating probe (denoted with numbers in Figure 3) are the low TRLs of AI technologies (1). Undeveloped AI systems constrain the scale of a self-replicating probe to a small-scale harvesting/manufacturing system manageable from Earth (Section 4.1). The implications of a small-scale system limit the replication module to manufacturing technologies that can manufacture a wide variety of components and use material resources efficiently, additive manufacturing technologies for µ-gravity with currently low TRL (2). The low TRL of additive manufacturing technologies and AI require every module to be as simple as possible in order to be replicated. For this reason, resource harvesting from large celestial bodies with strong gravity fields and atmospheres was discarded. Resource harvesting from large moons and planets would require copious amounts of energy, propellant and an additional propulsion system for take-off and land procedures. Moreover, mechanical stresses endured during a launch procedure would impose more stringent constraint on the self-manufactured components, which would require more sophisticated manufacturing systems. Limiting resource harvesting activities to asteroids resources (3), limits the harvesting methods to those functional in µ-gravity environments (4) (currently with low TRL), limiting the overall size of the probe (5). As the availability of radioactive materials in asteroids is limited (6) with 0.1 and 0.35 p.p.m.[62], the power generation system is restricted to operate with solar power technologies. Therefore, the overall mission is constrained to be performed inside the solar system.

The low TRL of additive manufacturing technologies for µ-gravity would also limit the minimum manufacturable additive manufacturing feature size, constraining shapes and sizes of additive manufacturing details of the various modules of the probe; component miniaturization would be restricted to the best available resolution of low TRL additive manufacturing machines for µ-gravity. On the other side, the size of the largest components of the probe, such as the antenna, are limited by the building capacities of the additive manufacturing system.

### 4.1. Bring or build

From the previous sections, it can be concluded that every system in the probe can be, theoretically, self-replicated. However, considering the current technology development of the replication and resource harvesting module, the manufacturing system necessary to fabricate every component on-board is itself difficult to manufacture.

Espera et al. [89], for example, present a comprehensible review of current additive manufacturing capabilities for electronic applications. However, additive manufacturing processes for solar arrays and electronic as well as optical equipment in general require large, heavy and complex additive manufacturing machines. Such machines would require large resource harvesting and sophisticated manufacturing systems similar to those proposed by Freitas [3] and which are limited by currently underdeveloped AI systems (Section 4.1) which makes them an unsuitable candidate for self-replication.

For these practical reasons, we propose that some components of the self-replicating probe are carried from Earth, a strategy previously proposed by Dunn et al. [90]. The idea is to establish a practical degree of partial self-replication, which can then be improved via further technological developments. Table 7 presents a summary of components on board of a self-replicating probe distinguishing between those that should be brought from Earth from those manufactured on board.

**Table 7. Manufacturing status of components on board the self-replicating probe**

| Module | Component | Status |
|---|---|---|
| Telemetry, tracking, command and instrument | Computers, chips, control systems, sensors, radios, general electronic components. | Bring |
| | Mechanical structures, connectors, antennas, amplifiers | Replicate |
| Power generation | General electronic components, solar cells | Bring |
| | Structures for solar arrays, connectors | Replicate |
| Replication | Mechanical components | Replicate |
| | Optic/laser components for LPB additive manufacturing machines | Bring |
| | General circuitry, power cables and general electric parts in components such as actuators | Bring |
| Resource harvesting | Mechanical components | Replicate |
| | General circuitry, power cables and general electric parts in components such as actuators | Bring |
| Propulsion | Mechanical components | Replicate |
| | General circuitry, power cables and general electric parts in components such as actuators | Bring |
| | Optic/laser components for some propulsion systems | Bring |
| General mechanical supports and brackets | Every component | Replicate |

## 4.2. Concept

The proposed concept for a self-replicating probe is a < 100 kg probe (without counting mass of stored components for replicated probes) with approximately 18 m$^2$ of solar panels. Such a concept achieves replication of its mechanical components through metal clay additive manufacturing technology, as it is the additive manufacturing alternative with the lowest weight and energy consumption. The resource harvesting is performed with a robotic arm with a sample-spoon like end. Such an arm can be repurposed for assembling the manufactured components.

The rest of the module's specifications and other possible technology alternatives are detailed in Table 8.

**Table 8: Technology alternatives and parameters for self-replicating probe modules.**

| Module | Technology alternatives | Volume, weight, energy consumption |
|---|---|---|
| Telemetry, tracking, command and instrument | Flight computer, attitude control, radio and navigation equipment, Command and Data Handling board like in MarCo mission: on-board non-volatile storage, real-time clock, and cascaded watchdog system, and interfaces to all on-board subsystems. The software allows for uploadable sequences, storage and transmission of real-time and historical telemetry [30]. Radiators, heaters, temperature and environment sensors. Non-deployable reflectarray antenna. Batteries. | A space of a 6U CubeSat, 25kg, 17W [30] |
| Power generation | 18 m² [91,92] of Laminated perovskite solar cells. | |
| Propulsion | Laser ablation plasma: (<0.2) m/s | |
| | Vaporizing liquid: (0.2-16) m/s | |
| | Electrothermal (resistojets): (0.6-160) m/s | |
| Replication | Additive manufacturing with metal clay systems. The outside of M asteroids can provide metallic powder, C type asteroids can provide organic binder [90]. | (0.2 × 0.3 × 0.5)m, 5kg [64], 50W |
| | Laser powder bed additive manufacturing | (0.74 x 0.63 x 1) m, >100kg, 1kW [63,93] |
| Resource harvesting | Metallic resources: Robotic arm inspired on the Osiris-Rex TAGSAM system [33] and/or projectile firing like Hayabusa system [78] Water and volatiles: heater deep flute with perforated walls for melting and/or sublimating volatiles like in the Sniffer system [77]. | 18U CubeSat, 40kg 200W/hs [77] |

The implementation of LPB additive manufacturing technologies is feasible, however its weight and complexity (the need for a gas flow-assisted powder deposition, for instance) would render the replication process problematic. In addition, LPB additive manufacturing technologies have a larger energy consumption than the metal clay counterpart.

The electronic components of the telemetry, tracking, command and instrument module will not be self-manufactured and several copies of it would be carried from Earth to be transferred to probes manufactured on space. The solar array will be also brought from Earth in the shape of perovskite laminated solar array rolls [49].

The propulsion system of choice is based on resistojets (large ΔV in comparison with the other feasible options) combined with gravity assist maneuvers. Solar sails can also be implemented, although they must be carried from earth, as metal clay additive manufacturing technologies are not yet able to manufacture such thin material layers. In this context, the probe is expected to be able to replicate roughly 70% of its mass on board, with the mass breakdown shown in Table 9.

The replicated components also include some of the parts of the power (solar array structure) and propulsion module (structural components) are also replicated (~2 kg).

**Table 9: Mass breakdown of non-replicated and replicated spacecraft modules**

| Spacecraft modules | Replicated (yes/no) | Mass [kg] |
|---|---|---|
| Telemetry, tracking, command and instrument, power, propulsion | No | 23 |
| Replication, resource harvesting, elements of propulsion | Yes | 47 |

Figure 4 shows the configuration of the near-term self-replicating probe. The main spacecraft body consists of three sections (see Table 8). The resource harvesting module, replication module, and the back end. The back end includes the telemetry, tracking, command and instrumentation module, the component storage compartment, the propulsion and power module. The resource harvesting module is located at the front. The attached robotic arm has the purpose of collecting asteroid material, which is introduced into the resource entry opening. This module processes the asteroid material and the resulting metal clay is transferred to the replication module, where component manufacturing takes place. The finished components are then assembled via the robotic arm, and stored in the storage compartment. The resistojects in the propulsion module are located in two clusters on the top and bottom of the spacecraft as well as the remaining propulsion system elements, which are located inside the back compartment. All resistojets are used for attitude control, however, the resistojets facing into the back direction of the spacecraft are also used for moving between asteroids. The antenna is attached to the back plane, which eventually needs to accommodate the launch adapter, at least for the initial spacecraft, which is launched from Earth. The power subsystem with its solar array, are mounted on the back end.

The replicated components are mostly based on plate-like structures, which are then assembled via the robotic arm. The central section of the spacecraft would easily fit into a door frame and would be roughly the size of a very large suitcase.

Figure 5 shows a view of the front of the spacecraft in its entirety. The large size of the solar panels compared to the central section of the spacecraft can be seen. In Figure 6, an artistic representation of the self-replicating probe in operation, is presented.

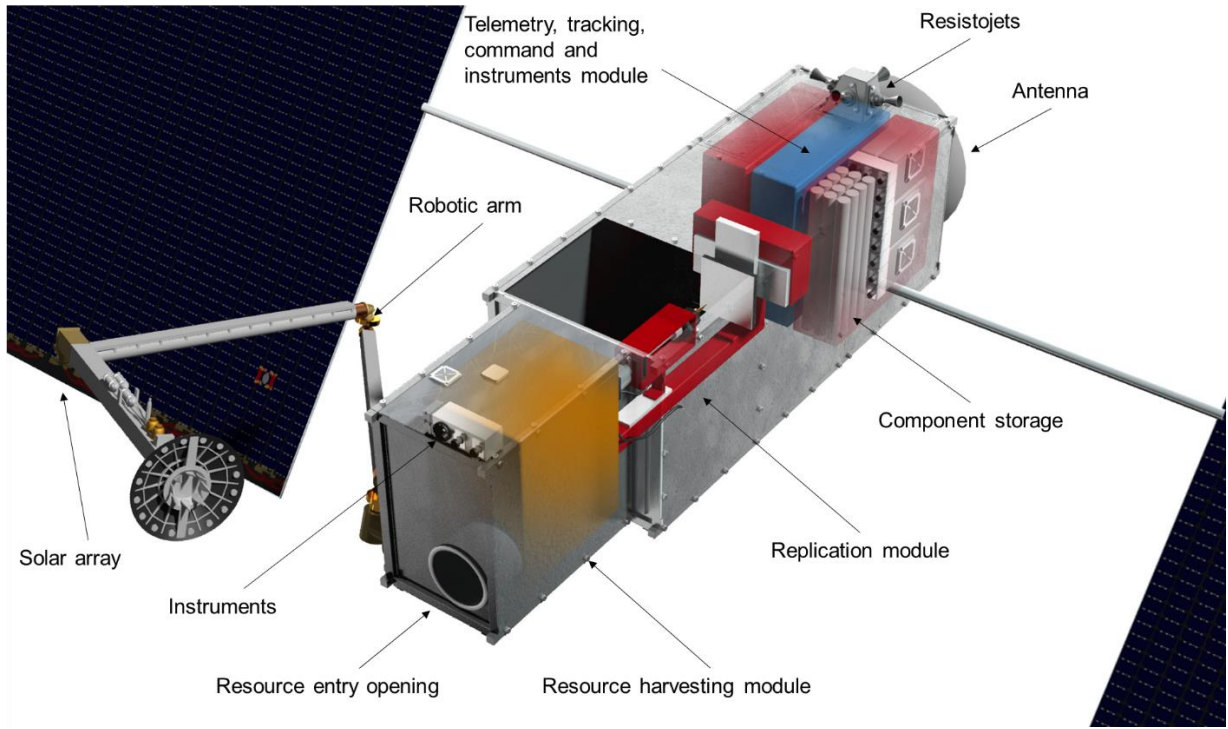

**Figure 4. Configuration of a near-term self-replicating probe, based on the proposed concept (Credit: Adrian Mann)**

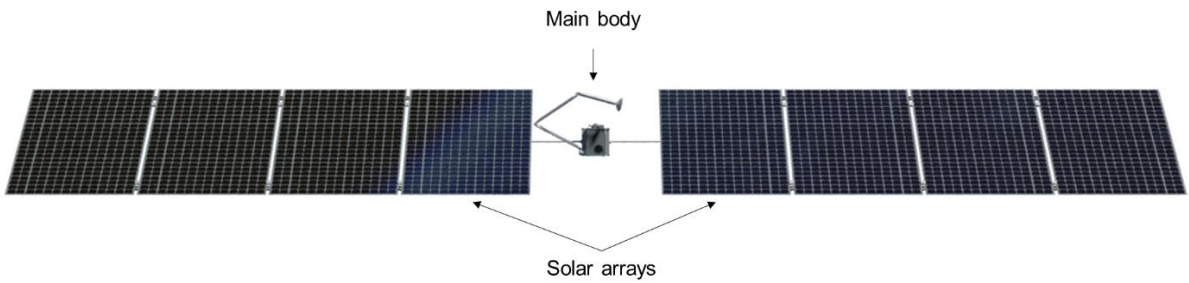

**Figure 5. Front view of a near-term self-replicating probe (Credit: Adrian Mann)**

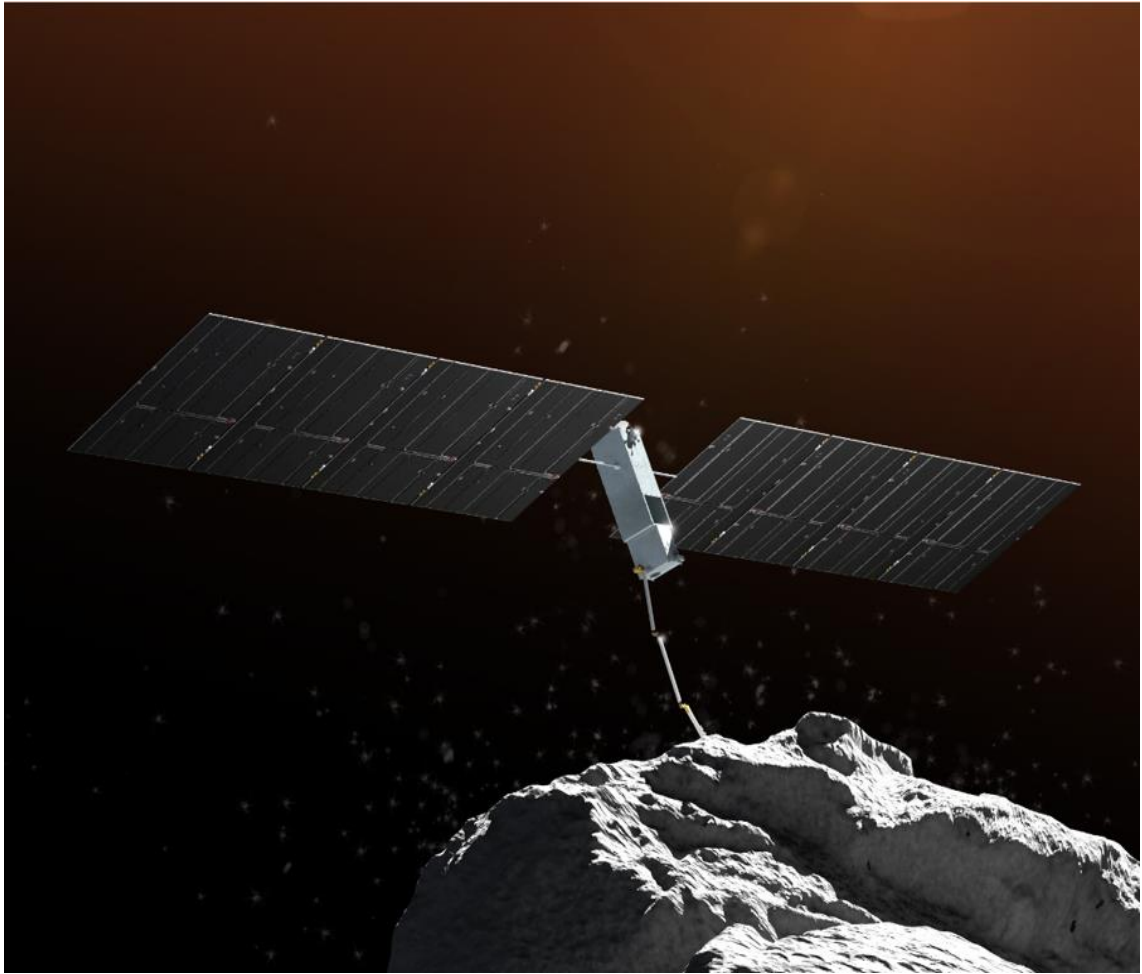

**Figure 6. Artistic representation of the proposed self-replicating probe in operation (Credit: Adrian Mann)**

## 5. Cost Comparison

In the following, we demonstrate that launching a solar system-based partially self-replicating probe able to replicate a finite number of times can still have economic benefits and increase the efficiency of space exploration. Partial self-replication means that replication cannot be sustained indefinitely, as the initially supplied non-replicated components will run out. Nevertheless, substantial economic benefits from launching self-replicating probes are expected, as suggested by Figure 7.

In Figure 7, a sample comparison of launch costs of a self-replicating probe carrying material for *n* replication cycles and launch costs of launching *n* conventional probes (without replication and resource harvesting modules), is presented. We focus on launch costs and do not consider other cost factors, such as development cost, as they would strongly depend on the compared spacecraft architectures. Launch costs are calculated considering the utilization of a launcher such as the Atlas V, with an approximate specific launch cost of 14 k$/kg. The breakeven points of different

replication percentages are indicated via stars. A conventional probe is lighter, as no modules for self-replication are required. For simplicity, we take the same spacecraft but without the respective modules. This results in 25 kg, using 23 kg from Table 9 for all non-replicated components and adding further 2 kg for power and propulsion components, which are also not replicated. This can be seen for the launch cost of the first unit, as the conventional probe is substantially lighter. However, with higher numbers of probes, partial self-replication plays out, as less mass needs to be launched for subsequent probes. Hence, the (partial) self-replication capability is only advantageous, once a certain number of probes is surpassed.

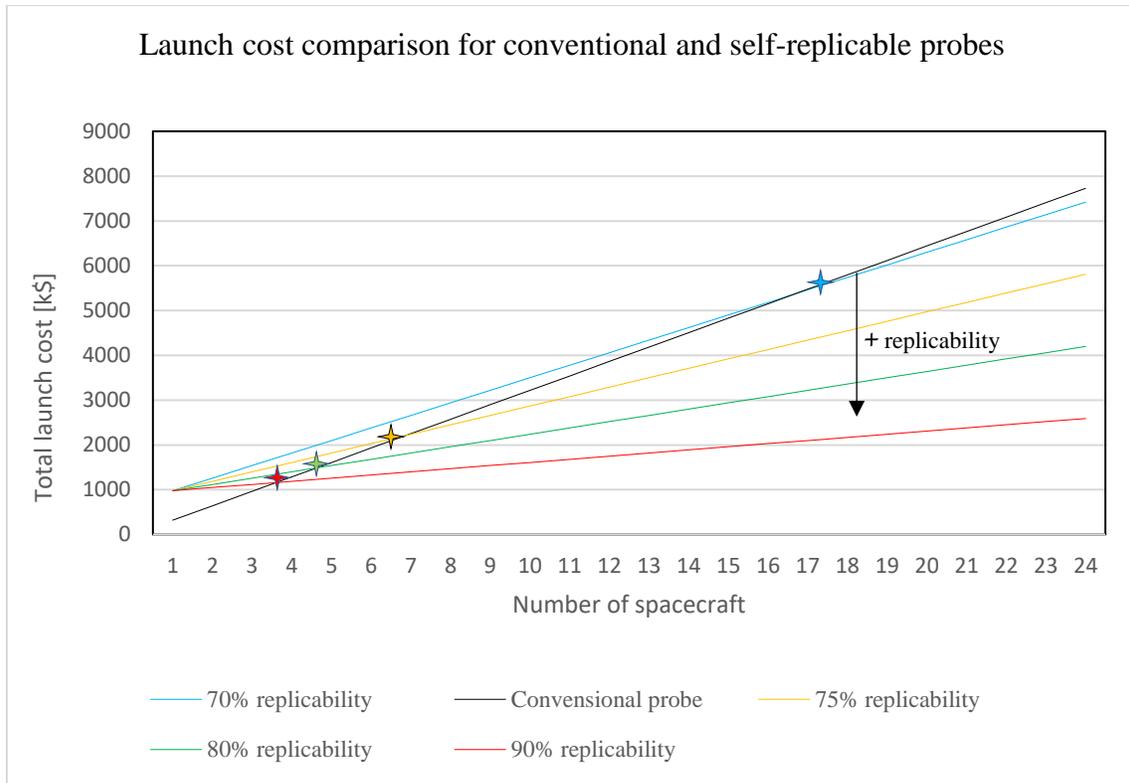

**Figure 7. Launch costs comparison between a self-replicating and a conventional probe.**

The current probe design is estimated to reach 70% replication of its mass, which renders the probe economically convenient after approximately 17 replications. Increasing the replication capabilities, results in self-replication probes that are cheaper to launch after 7, 5 and 4 replication cycles for 75%, 80% and 90%, respectively. It can be seen that an increase from 70 to 75% would lead to a drastic decrease in cost but additional improvements to 80 and 90% have diminishing returns on cost. The main reason is that the hard-to-replicate components, e.g. microchips also tend to have a low mass.

Even if current self-replication capabilities limit the economic benefits of a self-replicating probe, such probe could be manufactured in the next 10 years and fulfil objectives related to technical demonstration (collect data about probe capabilities) and space exploration. Future work might reveal a different picture, when all cost factors are taken into account. For example, self-replicating probes might cost more to develop than conventional probes, thereby diminishing the cost advantage.

## 6. Technology Roadmap

Progressively advancing TRLs of the three main limiting technologies, AI, additive manufacturing and resource harvesting, will amplify the probe capabilities as presented in Figure 8. Advances in AI could enable the development of large-scale self-replicating systems, as those proposed by Bond and Martin [24] or Freitas [3], where probes of hundreds of tons would carry a plethora of manufacturing systems able to manufacture each component on board. On the other side, advancements in AI would enable self-replicating probes to successfully land and build resource harvesting and manufacturing facilities on resource abundant planets and moons. In the long term, access to large celestial bodies can provide the required radioactive material for fueling the nuclear power system of a probe, enabling it to leave the solar system. Another convenient AI technology for outer solar system exploration are digital twins. Digital twin systems can monitor the status of the probe and its components and perform repair or preventive maintenance activities, a desirable function for interstellar exploration [2].

Further developing additive manufacturing technologies would enable the miniaturization of additive manufacturing machines, making them lighter and perhaps less complex, and in need of less harvested material to be manufactured. Miniaturized additive manufacturing machines for manufacturing of electronics and perovskite SA would contribute to increasing self-replication capabilities while maintaining a low mass. At the same time, developing additive manufacturing technologies can reduce the minimum manufacturable feature size, which might allow for those miniaturized machines to be self-replicated in space. Advances in additive manufacturing would also be necessary to autonomously manufacture a propulsion system able to land and take off from a large celestial body.

Future developments of the resource harvesting module would allow for the recollection and processing of radioactive materials, such as for RTG systems as well as propellants such as hydrazine for propulsion systems with a higher Isp.

The development of advanced efficient fission reactors and advanced on-board material processing capabilities for isotopic enrichment would contribute to increase the probe capability to exit the solar system and cruise to another.

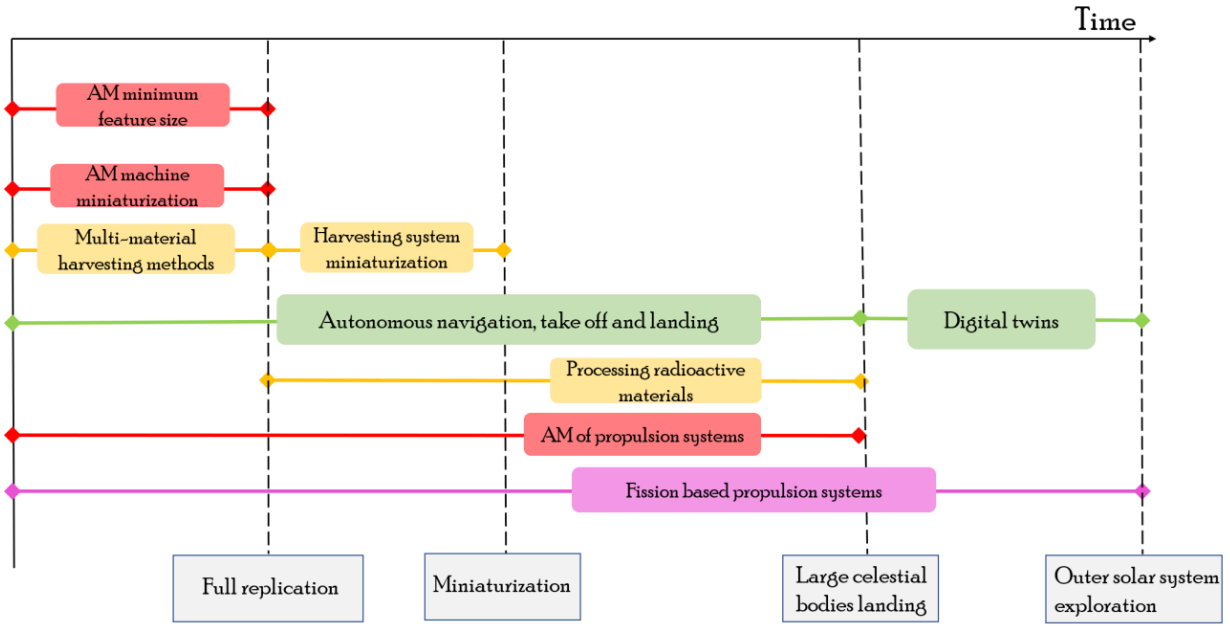

**Figure 8. Technology development areas that would extend the application and self-replicability of a self-replicating probe. AM: Additive manufacturing**

## 7. Discussion

The technology analysis performed in this article suggests that completely self-replicating probes are not feasible with current AI and additive manufacturing technologies. Due to the lack of development of those technologies, some components of the self-replicating probe, such as solar cells and electronics in general, need to be carried from Earth, constraining the number of replication cycles. Nevertheless, partial self-replication seems to be feasible and might provide significant cost advantages, compared to conventional probes, once a certain number of probes is required for a mission. Even if current self-replication capabilities limit the cost benefits of a self-replicating probe, such a probe could be manufactured in the next 10 years and fulfil objectives related to technical demonstration and space exploration.

Low technology maturity limits the operational domain of the probe to our solar system, as the probe is limited to harvest resources from asteroids, and is not able to harvest enough radioactive material to supply nuclear power sources, which would be required for missions in the outer solar system and interstellar missions.

A limitation of our study is that we sacrificed the level of detail of the self-replication process itself, in order to focus more on the required technologies, high-level feasibility, and rationale. Future work could bring this analysis together with a more detailed analysis of the self-replication process.

# 8. Conclusion

Self-replicating probes have the potential to exponentially increase the number of probes and thereby improve the efficiency of space exploration. Through a technology analysis, it was concluded that the current maturity of artificial intelligence, additive manufacturing and asteroid resource harvesting technologies hinder the development of a completely self-replicating probe. However, partially self-replicating probes are feasible. A small satellite-scale concept for a 70% self-replicable probe based on current and near-term technologies was proposed. Hence, we consider small-scale self-replicating probes to be feasible, and could be launched within the next 10 years, if microchips and other complex electronic components are brought with the initial probe and are not replicated. Such probes would still be an important technology demonstration and under certain conditions, be more cost-effective for exploration missions than non-replicating probes.

## Acknowledgements

We would like to thank Chris Corner, who founded the i4is von Neumann Project and laid the groundwork on which this research is based. He passed away in March 2019.

Technol. 4 (2019) 1900506.
[43] Asteroid Mining Corporation, Asteroid Mining Probe One (AMP-1) 2028, (n.d.). https://asteroidminingcorporation.co.uk/our-missions (accessed April 1, 2020).
[44] PlanetaryResources, Planetary Resources & 3D Systems Reveal First Ever 3D Printed Object From Asteroid Metals, (2016). http://www.planetaryresources.com/2016/01/planetary-resources-and-3d-systems-reveal-first-ever-3d-printed-object-from-asteroid-metals/ (accessed April 1, 2020).
[45] 3D Print Mixed Metals, Magnets and Open Materials: A Look at Emerging New Company Formalloy at RAPID 2016, (2016). https://3dprint.com/135396/formalloy-rapid-2016/ (accessed April 1, 2020).
[46] S. Marquant, Amazing 3D Printer Can Use 21 Different Metals Simultaneously, (2016). https://futurism.com/nvbots-introduces-technology-can-3d-print-using-21-different-metals-simultaneously (accessed April 1, 2020).
[47] DragonFly, Lights-out Digital Additive Manufacturing for Printed Electronics- Product catalog. Nano Dimension Ltd, (2019). https://www.nano-di.com/nano-dimension-brochures (accessed May 5, 2020).
[48] S.K. Karunakaran, G.M. Arumugam, W. Yang, S. Ge, S.N. Khan, X. Lin, G. Yang, Recent progress in inkjet-printed solar cells, J. Mater. Chem. A. 7 (2019) 13873–13902.
[49] R. Schmager, J. Roger, J.A. Schwenzer, F. Schackmar, T. Abzieher, M. Malekshahi Byranvand, B. Abdollahi Nejand, M. Worgull, B.S. Richards, U.W. Paetzold, Laminated Perovskite Photovoltaics: Enabling Novel Layer Combinations and Device Architectures, Adv. Funct. Mater. (2020) 1907481.
[50] J. Yoon, S. Jo, I.S. Chun, I. Jung, H.-S. Kim, M. Meitl, E. Menard, X. Li, J.J. Coleman, U. Paik, GaAs photovoltaics and optoelectronics using releasable multilayer epitaxial assemblies, Nature. 465 (2010) 329–333.
[51] F. Mathies, E.J.W. List-Kratochvil, E.L. Unger, Advances in Inkjet-Printed Metal Halide Perovskite Photovoltaic and Optoelectronic Devices, Energy Technol. 8 (2020) 1900991.
[52] B.J.G. de la Bat, R.T. Dobson, T.M. Harms, A.J. Bell, Simulation, manufacture and experimental validation of a novel single-acting free-piston Stirling engine electric generator, Appl. Energy. 263 (2020) 114585.
[53] C.C. Morrill, Apollo fuel cell system, in: Proc. 19th Ann. Power Sources Conf., 1965: pp. 38–41.
[54] M. Cifrain, K. Kordesch, Hydrogen/oxygen (air) fuel cells with alkaline electrolytes, Handb. Fuel Cells. (2010).
[55] A. Meurisse, J. Carpenter, Past, present and future rationale for space resource utilisation, Planet. Space Sci. 182 (2020) 104853.
[56] D. Buden, J.A. Angelo Jr, Nuclear energy-key to lunar development, in: Lunar Bases Sp. Act. 21st Century, 1985: p. 85.
[57] M. Kaya, V. Bozkurt, Thorium as a nuclear fuel, in: 18th Int. Min. Congr. Exhib. Turkey–IMCET, 2003: pp. 571–578.
[58] J.J. Hagerty, D.J. Lawrence, B.R. Hawke, L.R. Gaddis, Thorium abundances on the Aristarchus plateau: Insights into the composition of the Aristarchus pyroclastic glass deposits, J. Geophys. Res. Planets. 114 (2009).
[59] N. Yamashita, N. Hasebe, R.C. Reedy, S. Kobayashi, Y. Karouji, M. Hareyama, E. Shibamura, M. Kobayashi, O. Okudaira, C. d'Uston, Uranium on the Moon: Global distribution and U/Th ratio, Geophys. Res. Lett. 37 (2010).
[60] A.T. Bazilevskii, L.P. Moskaleva, O.S. Manvelian, I.A. Surkov, Evaluation of the thorium and uranium contents of Martian surface rock-A new interpretation of Mars-5 gamma-spectroscopy measurements, Geokhimiia. (1981) 10–16.
[61] G. Crozaz, P. Pellas, M. Bourot-Denise, S.M. de Chazal, C. Fiéni, L.L. Lundberg, E. Zinner, Plutonium, uranium and rare earths in the phosphates of ordinary chondrites—The quest for a chronometer, Earth Planet. Sci. Lett. 93 (1989) 157–169.
[62] H.C. Urey, The cosmic abundances of potassium, uranium, and thorium and the heat balances of